\documentclass[twocolumn]{aastex62}

\begin{document}
\title{\bf \large Constraining the Gravitational Lensing of $z\gtrsim6$ Quasars from their Proximity Zones}
\shorttitle{Lensing from $z\gtrsim6$ Proximity Zones}
\shortauthors{F. B. Davies et al.}

\author[0000-0003-0821-3644]{Frederick B. Davies}
\affiliation{Lawrence Berkeley National Laboratory, 1 Cyclotron Rd, Berkeley, CA 94720, USA}

\author[0000-0002-7633-431X]{Feige Wang}
\altaffiliation{NHFP Hubble Fellow}
\affil{Steward Observatory, University of Arizona, 933 North Cherry Avenue, Tucson, AZ 85721, USA}

\author[0000-0003-2895-6218]{Anna-Christina Eilers}
\altaffiliation{NHFP Hubble Fellow}
\affiliation{MIT Kavli Institute for Astrophysics and Space Research, 77 Massachusetts Ave., Cambridge, MA 02139, USA} 

\author[0000-0002-7054-4332]{Joseph F. Hennawi}
\affiliation{Department of Physics, University of California, Santa Barbara, CA 93106-9530, USA}

\correspondingauthor{Frederick B. Davies}
\email{fdavies@lbl.gov}

\begin{abstract}
Since their discovery twenty years ago, the observed luminosity function of $z\gtrsim6$ quasars has been suspected to be biased by gravitational lensing. Apart from the recent discovery of UHS~J0439+1634 at $z\approx6.52$, no other strongly lensed $z\gtrsim6$ quasar has been conclusively identified. The hyperluminous $z\approx6.33$ quasar SDSS~J0100+2802, believed to host a supermassive black hole of $\sim10^{10} M_\odot$, has recently been claimed to be lensed by a factor of $\sim450$, which would negate both its extreme luminosity and black hole mass. However, its Ly$\alpha$-transparent proximity zone is the largest known at $z>6$, suggesting an intrinsically extreme ionizing luminosity. Here we show that the lensing hypothesis of $z\gtrsim6$ quasars can be quantitatively constrained by their proximity zones. We first show that our proximity zone analysis can recover the strongly lensed nature of UHS~J0439+1634, with an estimated magnification $\mu=28.0^{+18.4}_{-11.7}(^{+44.9}_{-18.3})$ at 68\% (95\%) credibility that is consistent with previously published lensing models. We then show that the large proximity zone of SDSS~J0100+2802 rules out lensing magnifications of $\mu>4.9$ at 95\% probability, and conclusively rule out the proposed $\mu>100$ scenario. Future proximity zone analyses of existing $z\gtrsim6$ quasar samples have the potential to identify promising strongly lensed candidates, constrain the distribution of $z\gtrsim6$ quasar lensing, and improve our knowledge of the shape of the intrinsic quasar luminosity function.
\end{abstract}

\section{Introduction}

Since the discovery of the first $z\sim6$ quasars \citep{Fan01}, the interpretation of their extremely high luminosity has been clouded by the potential for magnification by gravitational lensing \citep{WL02,Comerford02}. Understanding what fraction of the most luminous objects are lensed is crucial for constraining theories for how supermassive black holes formed, as the luminosities of the brightest objects may be vastly overestimated and bias our understanding of the quasar luminosity function \citep{Turner80}. While observational efforts to detect multiple images of $z>5$ quasars at 0\farcs1 resolution have been unsuccessful \citep{Richards06lens,McGreer14}, the recent discovery of the strongly lensed $z=6.5$ quasar UHS~J0439+1634 (\citealt{Fan19}; henceforth J0439+1634), and its closely separated lens galaxy, has lead to suggestions that a larger population of lensed quasars may have been missed due to color selection biases \citep{Fan19,PL19}. Small separation lenses may also be missed due to cuts on source morphology which are commonly applied to select high-redshift quasar candidates (e.g. \citealt{Richards02,Wang17}).

Recently, \citet{Fujimoto20} suggested that the brightest $z>6$ quasar known, SDSS~J0100+2802 (\citealt{Wu15}; henceforth J0100+2802) is strongly gravitationally lensed, based on re-analysis of high spatial resolution ALMA and HST imaging. They inferred a magnification of $\mu\sim450$, with a factor of a few uncertainty in the lensing model. This claim supported previous models (e.g. \citealt{WL02}) suggesting a substantial fraction of $z\gtrsim6$ quasars are strongly lensed \citep{PL20}. However, noted by both \citet{Fujimoto20} and \citet{PL20}, one property of J0100+2802 potentially defies the strongly lensed hypothesis: the extent of its proximity zone.

The ``proximity zone'' of a quasar is defined as the region where the quasar has substantially over-ionized the surrounding intergalactic medium (IGM) relative to the cosmic mean. The relative lack of neutral hydrogen in the vicinity of the quasar leads to an excess of transmission (or, equivalently, a deficit of absorption) in the Ly$\alpha$ forest blueward of rest-frame Ly$\alpha$ \citep{Bajtlik88}. The proximity zone size of $z\gtrsim6$ quasars is often quantified by ``$R_p$'', defined by \citet{Fan06} as the point at which the (observed-frame) 20\,{\AA}-smoothed Ly$\alpha$ transmission first drops below 10\%. While this definition is somewhat ad hoc, $R_p$ has been shown to be a valuable tool for constraining quasar lifetime \citep{Eilers18J1335}, and is sensitive to the size of ionized regions around reionization-epoch quasars \citep{Davies19}. Radiative transfer simulations of proximity zones show remarkable consistency with the observations across a wide range of quasar luminosity and redshift without parameter tuning \citep{Eilers17,Davies20a}. 

A small fraction of $z\gtrsim6$ quasars, however, show much smaller proximity zones than expected, which has been interpreted as a consequence of short quasar lifetimes compared to the photoionization timescale \citep[see e.g.][]{Khrykin16} of the IGM \citep{Eilers17,Eilers20}. Indeed, \citet{Eilers17} noted that J0100+2802 has a smaller proximity zone than expected given its observed brightness, and suggested that its lifetime may be short, $t_{\rm q}\lesssim 10^5$ yr. However, given the dependence of $R_p$ on the \emph{intrinsic} quasar luminosity ($R_p\propto L^{\sim1/2}$ at low IGM neutral fraction, e.g. \citealt{BH07,Eilers17,Davies20a}), another possibility is that the quasar luminosity inferred from its apparent magnitude has been overestimated due to magnification by gravitational lensing. The one $z>6$ quasar which has been unambiguously confirmed to be strongly magnified by a foreground lens, UHS J0439+1634 at $z=6.52$ (\citealt{Fan19}; henceforth J0439+1634), was found to have a small proximity zone compared to the expectation for its observed magnitude, providing additional evidence for its lensed nature. This observation suggests that proximity zones can be used to test the strongly lensed hypothesis for $z>6$ quasars in general.

\citet{HC02} were the first to show that the size of $z\gtrsim6$ quasar proximity zones could be used to constrain their intrinsic luminosity, using the connection between the proximity zone size and the size of the ionized bubble carved out by a quasar in a neutral IGM. 
However, their method required the assumption that the IGM was mostly neutral at $z\sim6.3$, unlikely given recent constraints on the reionization history at $z\gtrsim7$ (e.g. \citealt{Planck18,Greig17b,Davies18b,Mason18,Wang20}), and suffered from complete degeneracy with quasar lifetime. The general idea has existed for quite some time: the seminal work of \citet{Bajtlik88} suggested that one could constrain the lensing hypothesis via the proximity effect (see also \citealt{Bechtold94}).

The quasar Q1208+1011 at $z\approx3.8$, once the highest redshift quasar known \citep{Hazard86}, showed indications of being strongly lensed via the detection of multiple images at small angular separation \citep{Magain92,Bahcall92img}. While subsequent spectroscopic investigation showed that both sources were quasars at the same redshift, their broad emission lines were not perfectly identical \citep{Bahcall92spec}, and no lensing galaxy has been conclusively identified \citep{Lehar00}, leaving the true nature of the source as a lingering question mark.
There has been a long standing debate \citep{Mortlock99} about whether sources like Q1208+1011 are lenses vs. binary quasars \citep{Hennawi06,Hennawi10}, but \citet{Giallongo99} argued that the lensing hypothesis for Q1208+1011 is strongly preferred from the relative lack of proximity effect in its spectrum, demonstrating the utility of the proximity effect for identifying gravitational lenses.

\begin{figure*}
\begin{center}
\resizebox{17.5cm}{!}{\includegraphics[trim={1.0em 1em 1.0em 1em},clip]{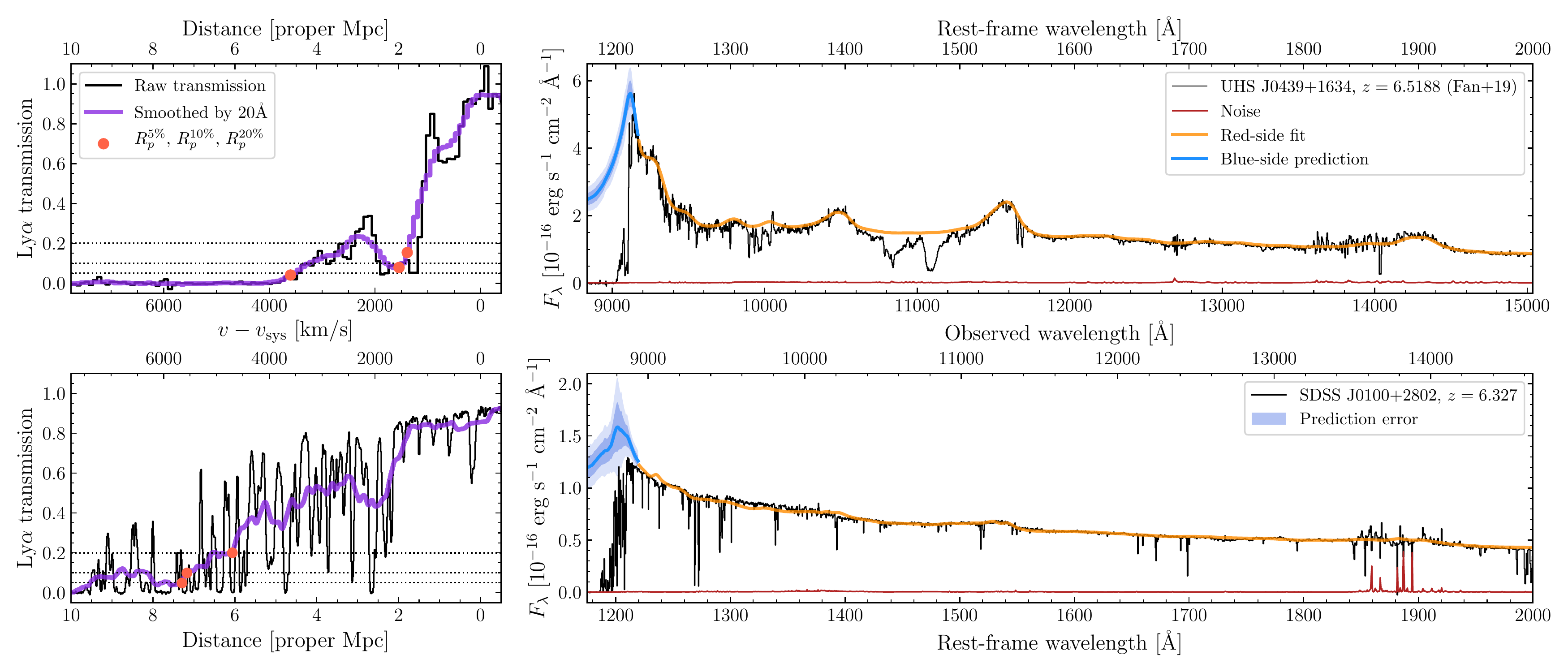}}
\end{center}
\caption{Top right: Keck/LRIS + Gemini/GNIRS spectrum of J0439+1634 from \citet{Fan19} (black) and its corresponding red-side PCA fit (orange) and blue-side prediction (blue). The dark and light shaded regions show the 1- and 2-$\sigma$ prediction error. Top left: Ly$\alpha$ transmission spectrum of J0439+1634 in the proximity zone (black), the 20\,{\AA}-smoothed spectrum (purple), and the $R_p$ values (dots) corresponding to the thresholds shown by the dotted lines. Bottom: The same for the VLT/X-SHOOTER spectrum of J0100+2802.}
\label{fig:spec}
\end{figure*}

In this Letter we show how proximity zone measurements can constrain the lensing magnification of $z\gtrsim6$ quasars using a suite of 1D ionizing radiative transfer simulations. First, we carefully re-measure the proximity zone sizes of J0100+2802 and J0439+1634 by reconstructing the unabsorbed quasar continua. We then demonstrate that the proximity zone of J0439+1634 prefers a strongly lensed solution, and infer magnification consistent with the lens models from \citet{Fan19}. Finally, we apply the same analysis to J0100+2802 and demonstrate that it cannot be strongly lensed.

In this work, we assume a $\Lambda$CDM cosmology with $h=0.685$, $\Omega_m=0.3$, $\Omega_\Lambda=0.7$ and $\Omega_b=0.048$.

\section{Data Analysis \& Proximity Zone Measurements} \label{sec:data}

We use the Gemini/GNIRS + Keck/LRIS spectrum of J0439+1634 from \citet{Fan19}. While J0439+1634 exhibits broad absorption line (BAL) features, the velocity range of the absorption does not influence the proximity zone. We correct for the foreground contamination by the lensing galaxy inside the proximity zone by subtracting the mean observed flux ($F_{\rm lens} \approx 3.8\times10^{-18}$ erg~s$^{-1}$~cm$^{-2}$~{\AA}$^{-1}$) just beyond the end of the transmission spikes in the proximity zone (8850\,{\AA} to 9000\,{\AA}) from the spectrum.

For J0100+2802, we reduced 2h of VLT/X-SHOOTER data acquired from the ESO archive\footnote{\url{http://archive.eso.org/eso/eso_archive_main.html}}. The observations were taken in August 2016 with exceptional seeing conditions under the program ID 096.A-0095 using the 0\farcs9 slit in the VIS arm and the 0\farcs6 slit in the NIR arm. Data reduction was performed applying standard techniques, including optimal extraction following \citet{Horne86}, using the open-source spectroscopic data reduction package PypeIt \citep{PypeItDOI,PypeItPaper}, which also includes the routines used for flux calibration, telluric correction, and optimal stacking of spectra. The extracted 1D spectra were flux calibrated using the standard star EG 274. Telluric correction was performed on each exposure by jointly fitting a PCA model of the quasar spectrum with a telluric model from a grid of Line-By-Line Radiative Transfer Model calculations (\texttt{LBLRTM}\footnote{\url{http://rtweb.aer.com/lblrtm.html}}; \citealt{Clough05}) produced using the python-based \texttt{TelFit} interface \citep{Gullikson14}. Finally, the fluxed and telluric corrected 1D spectra from each exposure were then stacked.

To model the intrinsic quasar spectra inside of their proximity zones, we use the principal component analysis (PCA) method of \citet{Davies18a}, as customized for post-reionization quasar proximity zones in \citet{Eilers20}. We briefly summarize the construction of the PCA here, but we point the interested reader to \citet{Davies18a} for details. A sample of 12,764 quasars were selected from the SDSS/BOSS DR12Q quasar catalog \citep{Paris17} at $2.09 < z < 2.51$ with signal-to-noise ratio greater than 7 at rest-frame 1290\,{\AA}. The spectra were queried and accessed via the \texttt{IGMSpec} database \citep{Prochaska17}. Each spectrum was normalized at 1290$\pm$2.5\,{\AA} and smoothed with an automated spline fitting routine from \citep{Dall'Aglio08}. Ten red-side (1220--2900\,{\AA}) and six blue-side (1175--1220\,{\AA}) PCA components were decomposed from nearest-neighbor stacks of the smoothed continua. The red coefficients for each quasar spectrum were then fit jointly with a degree of freedom in the redshift direction, and then the blue coefficients were fit in that ``PCA redshift'' frame. Finally, we derived a linear relationship between the red and blue coefficients (i.e., a projection matrix, as in \citealt{Suzuki05,Paris11}) via a linear least-squares solver, allowing us to predict the blue side spectrum from a fit to the red side.

In Figure~\ref{fig:spec}, we show the red-side PCA fits, blue-side predictions, and the Ly$\alpha$ transmission spectra in the proximity zones of J0439+1634 (top) and J0100+2802 (bottom). We measured $R_p$ from the transmission spectra by first convolving the spectrum with a 20{\AA} (observed-frame) boxcar filter, and then determining the distance to the first pixel\footnote{Here we differ very slightly from \citet{Eilers17}, who used the third consecutive pixel in their definition of $R_p$.} where the smoothed spectrum falls below 10\% (e.g. \citealt{Fan06}). We adopt quasar systemic redshifts from host galaxy [\ion{C}{2}] emission: $z=6.5188$ for J0439+1634 \citep{Yang19}, and $z=6.3270$ for J0100+2802 \citep{Wang19}.

Anticipating the degeneracy between lensing magnification and quasar lifetime, we measure additional ``$R_p$'' values at thresholds of 5\% and 20\%. For clarity, we will refer to these $R_p$ definitions with a superscript denoting the threshold: $R_p^{5\%}$, $R_p^{10\%}$, and $R_p^{20\%}$, where $R_p^{10\%}$ is the ``traditional'' $R_p$. For J0439+1634 we measure \{$R_p^{5\%}$,$R_p^{10\%}$,$R_p^{20\%}$\} = \{4.64, 1.99, 1.79\} proper Mpc, and for J0100+2802 we measure \{$R_p^{5\%}$,$R_p^{10\%}$,$R_p^{20\%}$\} = \{7.29, 7.17, 6.06\} proper Mpc. While our $R_p^{10\%}$ for J0100+2802 is nearly identical to the $R_p=7.14$ proper Mpc from \citet{Eilers17}, we note that our $R_p^{10\%}$ for J0439+1634 is substantially smaller than the $R_p=3.61\pm0.15$ proper Mpc from \citet{Fan19} due to our different continuum model and our subtraction of the flux from the foreground lens galaxy.

\begin{figure*}[ht]
\begin{center}
\resizebox{8cm}{!}{\includegraphics[trim={1.0em 1em 1.0em 1em},clip]{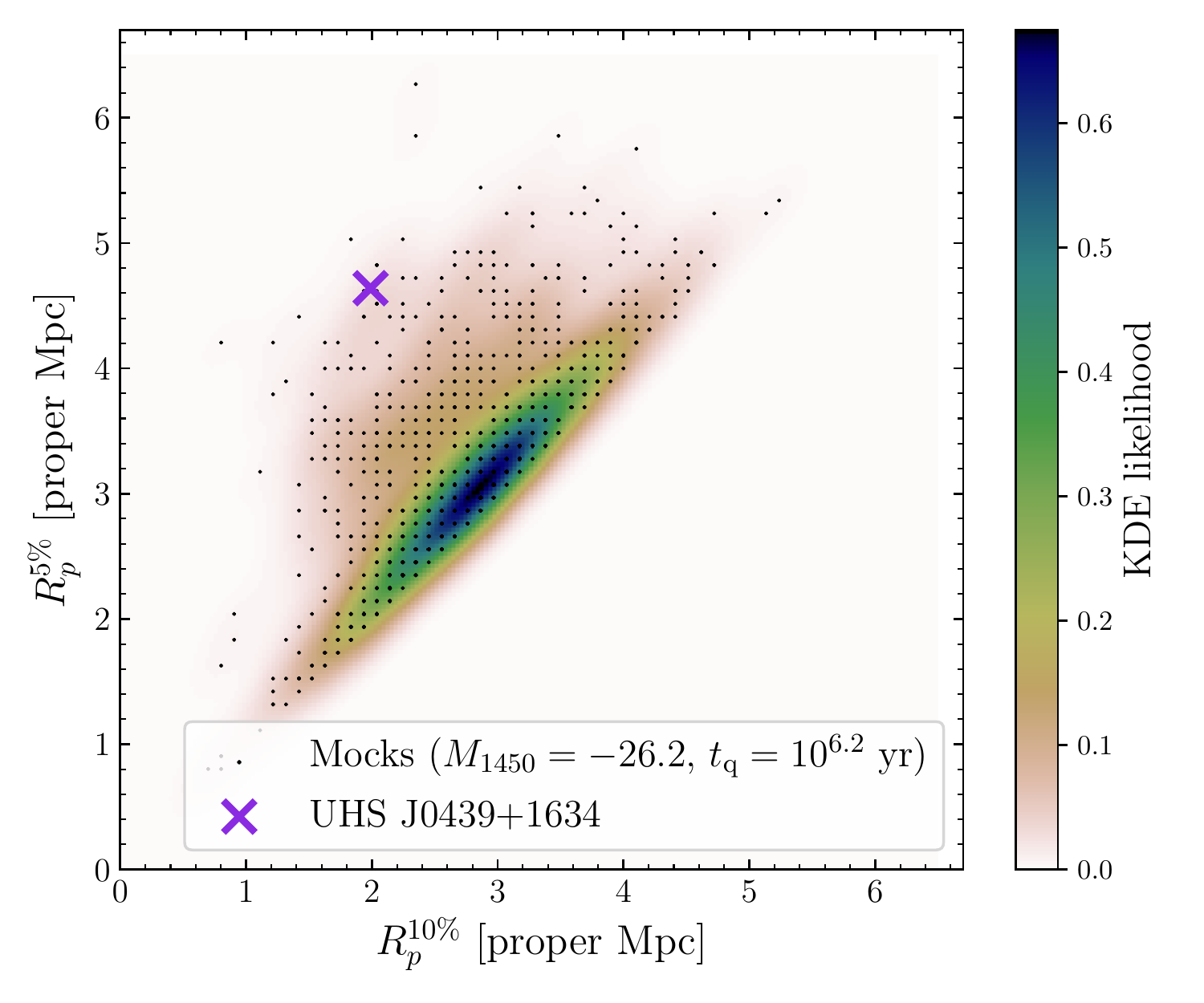}}
\resizebox{8cm}{!}{\includegraphics[trim={1.0em 1em 1.0em 1em},clip]{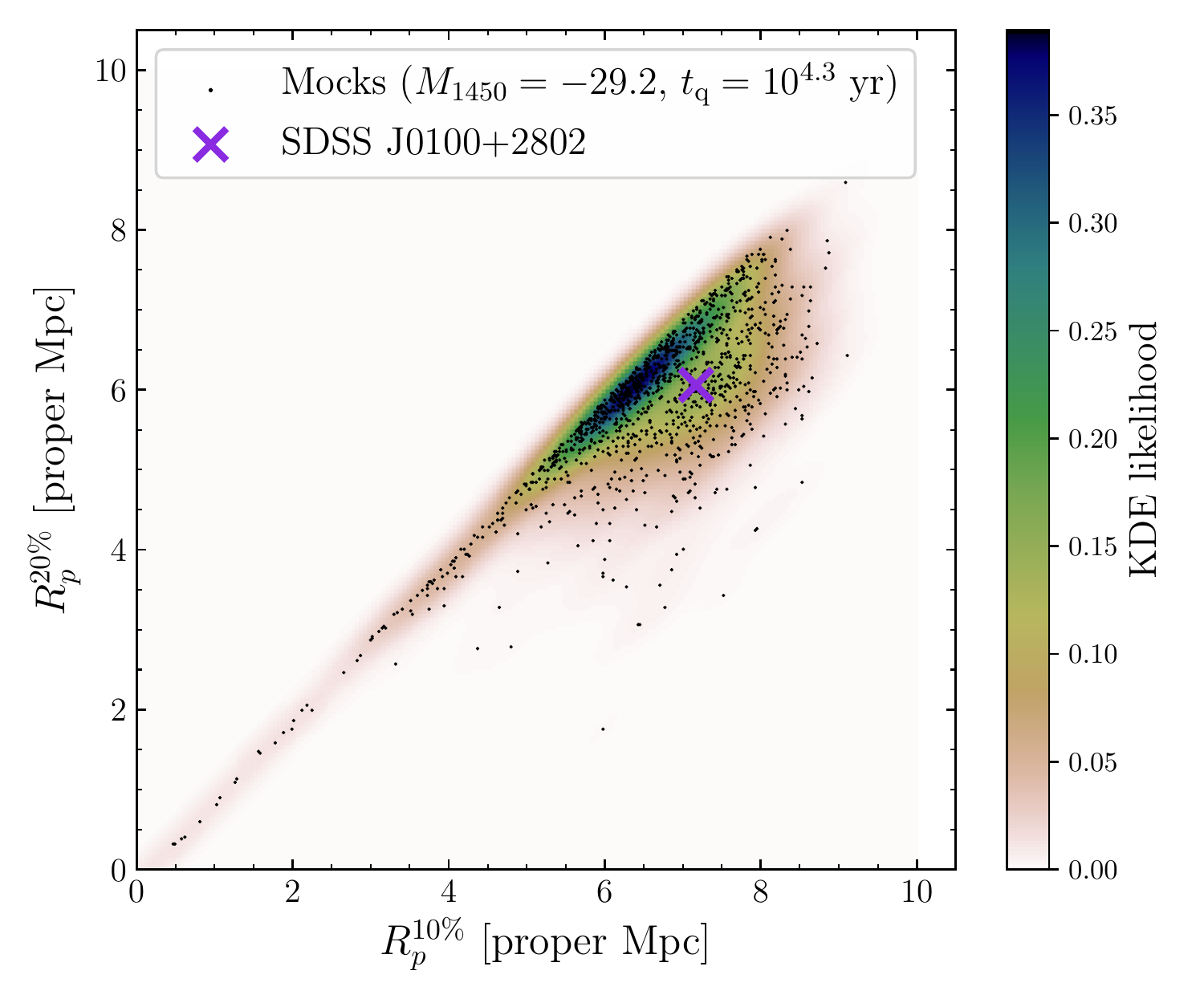}}
\end{center}
\caption{Left: Joint distribution of $R_p^{10\%}$ and $R_p^{5\%}$ from the $M_{1450}=-26.2$, $t_{\rm q}=10^{6.2}$ yr models in the J0439+1634 simulation suite (black points). The grid-like spacing of the $R_p$ values comes from our forward modeling of the coarse $\approx80$\,km/s pixel scale of the Keck/LRIS spectrum. Right: Joint distribution of $R_p^{10\%}$ and $R_p^{20\%}$ from the $M_{1450}=-29.2$, $t_{\rm q}=10^{4.3}$ yr models in the J0100+2802 simulation suite. The color shading in both panels represents a 2D KDE of the joint distributions, and the purple X marks the respective observed values of $R_p$.}
\label{fig:kde_ex}
\end{figure*}

\section{Proximity Zone Simulations} \label{sec:sims}

The size of the quasar proximity zone, as quantified by $R_p$, depends on the degree to which the quasar can ionize the IGM along the line of sight, and thus depends on the \emph{intrinsic} ionizing luminosity of the quasar (e.g. \citealt{BH07,Davies20a}). The quantitative connection between the two, however, depends on the considerable intrinsic scatter from IGM density fluctuations and time-evolution which can only be fully characterized via detailed simulations \citep{Eilers17,Davies20a}.

We simulate quasar proximity zones using the 1D ionizing radiative transfer code from \citet{Davies16} with minor updates described in \citet{Davies18b}. The code computes the time-dependent photoionization of hydrogen and helium in the presence of a quasar's ionizing radiation, and solves for the resulting thermal evolution including relevant primordial (i.e. metal-free) heating and cooling processes. For the density, initial temperature, and line-of-sight velocity along each simulated sightline, we draw 1200 skewers from the $z=6.5$ snapshot of a \texttt{Nyx} hydrodynamical simulation \citep{Almgren13,Lukic15} 100 comoving Mpc$/h$ on a side with $4096^3$ baryon cells and dark matter particles. Skewers were chosen to start from the locations of the 200 most massive dark matter halos ($M_h\gtrsim4\times10^{11}$\,M$_\odot$) and extend along the $\pm$ x,y,z coordinate directions.

We run simulations for each quasar on a grid of ionizing luminosities derived from a grid of effective quasar absolute magnitudes $M_{1450}^{\rm eff}$ from $-22$ to $-30$ in steps of $\Delta M=0.2$. We compute the ionizing luminosity of each model assuming the \citet{Lusso15} quasar spectral energy distribution (SED) to convert from $L_\nu(1450{\rm \AA})$ to $L_\nu(912{\rm \AA})$, and extrapolate to higher frequencies assuming $L_\nu \propto \nu^{-1.7}$. We then compute Ly$\alpha$ transmission spectra in steps of $\Delta\log{t_{\rm q}}=0.1$ from $t_{\rm q}=10^3$ to $10^8$ yr. Our suite of proximity zone models for each quasar thus consists of $41\times 51\times 1200$ transmission spectra.

We measure the proximity zone sizes of model spectra in the same way as the observed spectra, i.e. the distance at which 20\,{\AA}-smoothed transmission first drops below 5\%, 10\%, and 20\% (\{$R_p^{5\%}$,$R_p^{10\%}$,$R_p^{20\%}$\}). We modify each model spectrum with a draw from a multivariate Gaussian approximation to model the covariant error in the error in the PCA-predicted continuum following \citet{Davies18b}, although this has a very small effect on our analysis (see also \citealt{Eilers20}).

\section{Constraints on Lensing Magnification From Proximity Zones} \label{sec:constrain}

We specify the quasar parameters in terms of $M_{1450}^{\rm eff}$, a proxy for ionizing luminosity through the \citet{Lusso15} SED, and the quasar lifetime $t_{\rm q}$. We explore two different options for proximity zone summary statistics: 1. the standard proximity zone size $R_p=R_p^{10\%}$, and 2. a joint statistic consisting of the three proximity zone sizes with different flux thresholds $\{R_p^{5\%},R_p^{10\%},R_p^{20\%}\}$. At each parameter pair in the grid, we approximate the likelihood function with kernel density estimation (KDE) similar to \citet{Khrykin19}. 
We can then write the likelihood of a set of $R_p$ values $\vec{R}_p$ as
\begin{equation}
L(\vec{R}_p|M_{1450}^{\rm eff},t_{\rm q}) = p_{\rm KDE}(\vec{R}_p|M_{1450}^{\rm eff},t_{\rm q}),
\end{equation}
where $p_{\rm KDE}$ is the KDE of $\vec{R}_p$ derived from the set of forward-modeled mock spectra.

In the left panel of Figure~\ref{fig:kde_ex} we show an example of the KDE procedure in the 2D space of $R_p^{10\%}$ and $R_p^{5\%}$ for a model with $M_{1450}^{\rm eff}=-26.2$ and $t_{\rm q}=10^{6.2}$\,yr from the J0439+1634 model grid, where the points represent individual model skewers, the shading shows the KDE likelihood, and the purple X shows the J0439+1634 measurements. In the right panel of Figure~\ref{fig:kde_ex} we show a KDE evaluated in the space of $R_p^{10\%}$ and $R_p^{20\%}$ for a model with $M_{1450}^{\rm eff}=-29.2$ and $t_{\rm q}=10^{4.3}$\,yr from the J0100+2802 grid compared to the J0100+2802 $R_p$ measurements. The ($M_{1450}$,$t_{\rm q}$) models shown in Figure~\ref{fig:kde_ex} were chosen to have relatively high likelihood for the $\{R_p^{5\%},R_p^{10\%},R_p^{20\%}\}$ of each quasar. Note that the observed combination of $R_p$ values for J0439+1634 is relatively rare in our model grid, leading to somewhat noisy likelihood values, but similar spectra do exist. Contrary to J0439+1634, the $R_p$ measurements of J0100+2802 lie in a high probability region of the likelihood. 

\begin{figure*}[ht]
\begin{center}
\resizebox{8cm}{!}{\includegraphics[trim={1.0em 1em 1.0em 1em},clip]{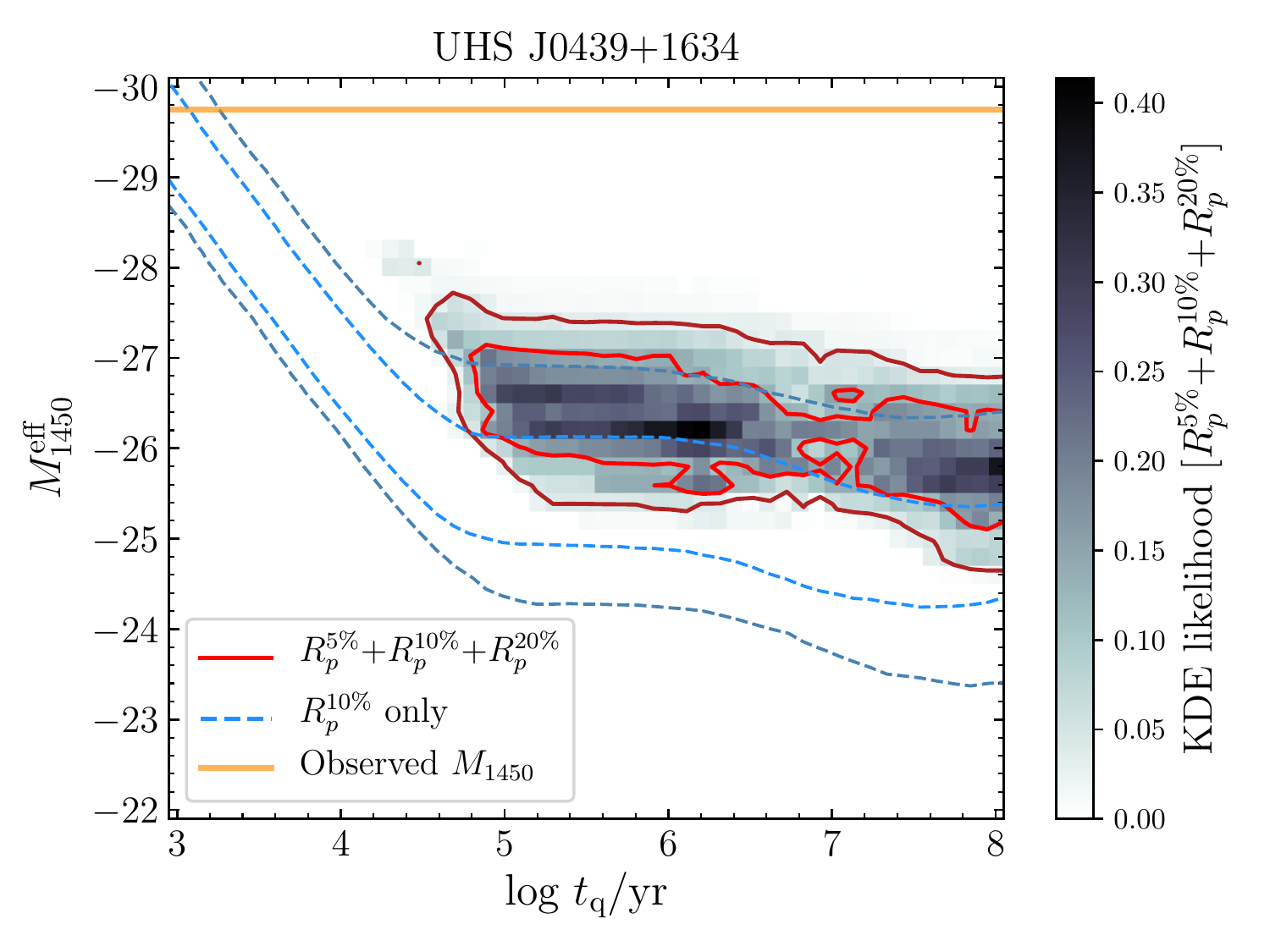}}
\resizebox{8cm}{!}{\includegraphics[trim={1.0em 1em 1.0em 1em},clip]{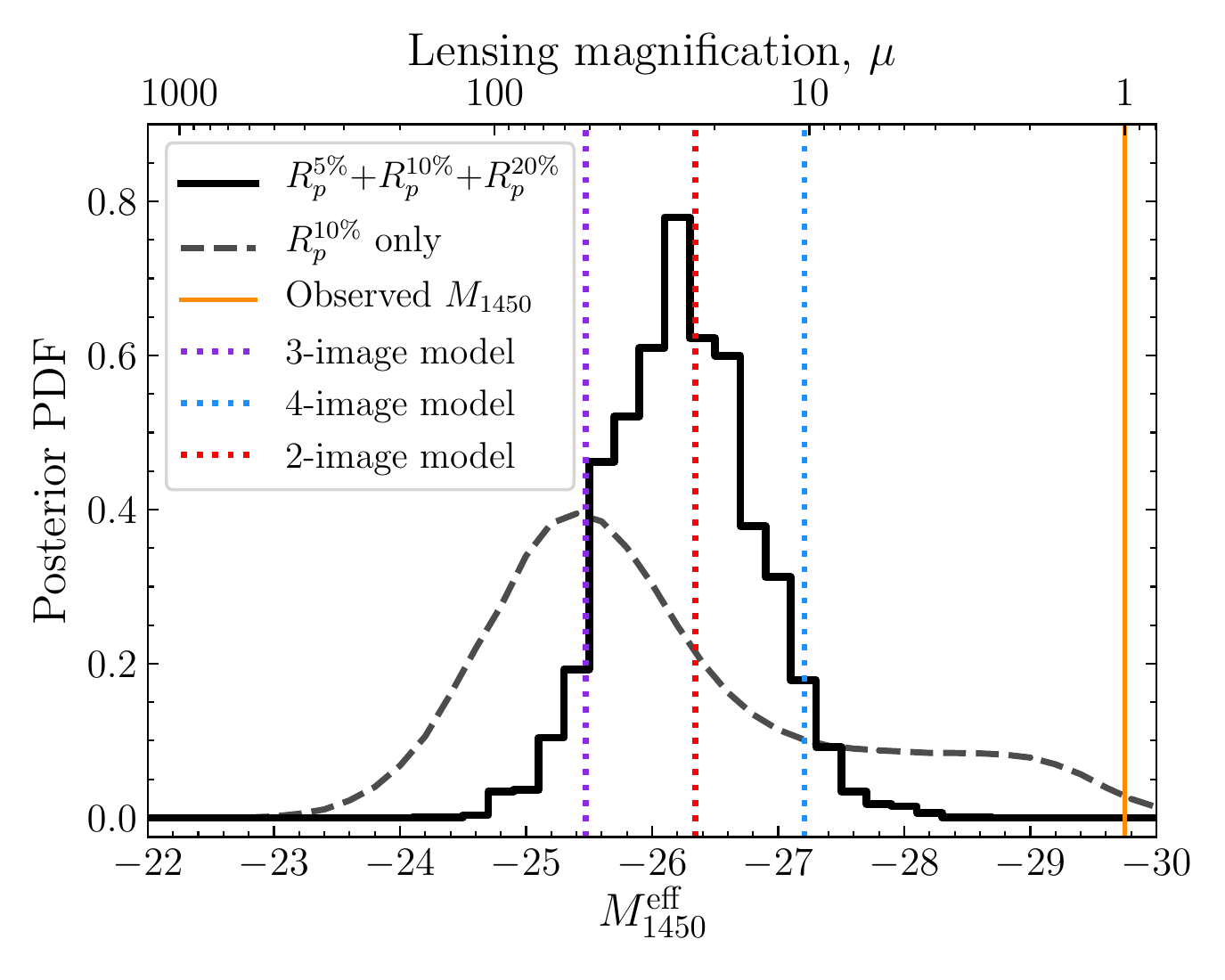}}
\end{center}
\caption{Left: Likelihoods of quasar lifetime $t_{\rm q}$ and effective absolute magnitude $M_{1450}^{\rm eff}$ for J0439+1634 using $\{R_p^{5\%},R_p^{10\%},R_p^{20\%}\}$ (shading plus red contours) and $R_p^{10\%}$ alone (blue dashed contours). The inner and outer contours enclose $68\%$ and $95\%$ of the likelihood, respectively. The orange line shows the observed $M_{1450}$ of J0439+1634. Right: Marginalized posterior PDF of $M_{1450}^{\rm eff}$ and lensing magnification $\mu$ (shown on the upper axis) using $\{R_p^{5\%},R_p^{10\%},R_p^{20\%}\}$ (solid) and $R_p^{10\%}$ alone (dashed). The orange line shows the observed $M_{1450}$, i.e. $\mu=1$, and the dotted lines show the $\mu$ values of the proposed lensing models in \citet{Fan19}.}
\label{fig:j0439}
\end{figure*}

\begin{figure*}[ht]
\begin{center}
\resizebox{8cm}{!}{\includegraphics[trim={1.0em 1em 1.0em 1em},clip]{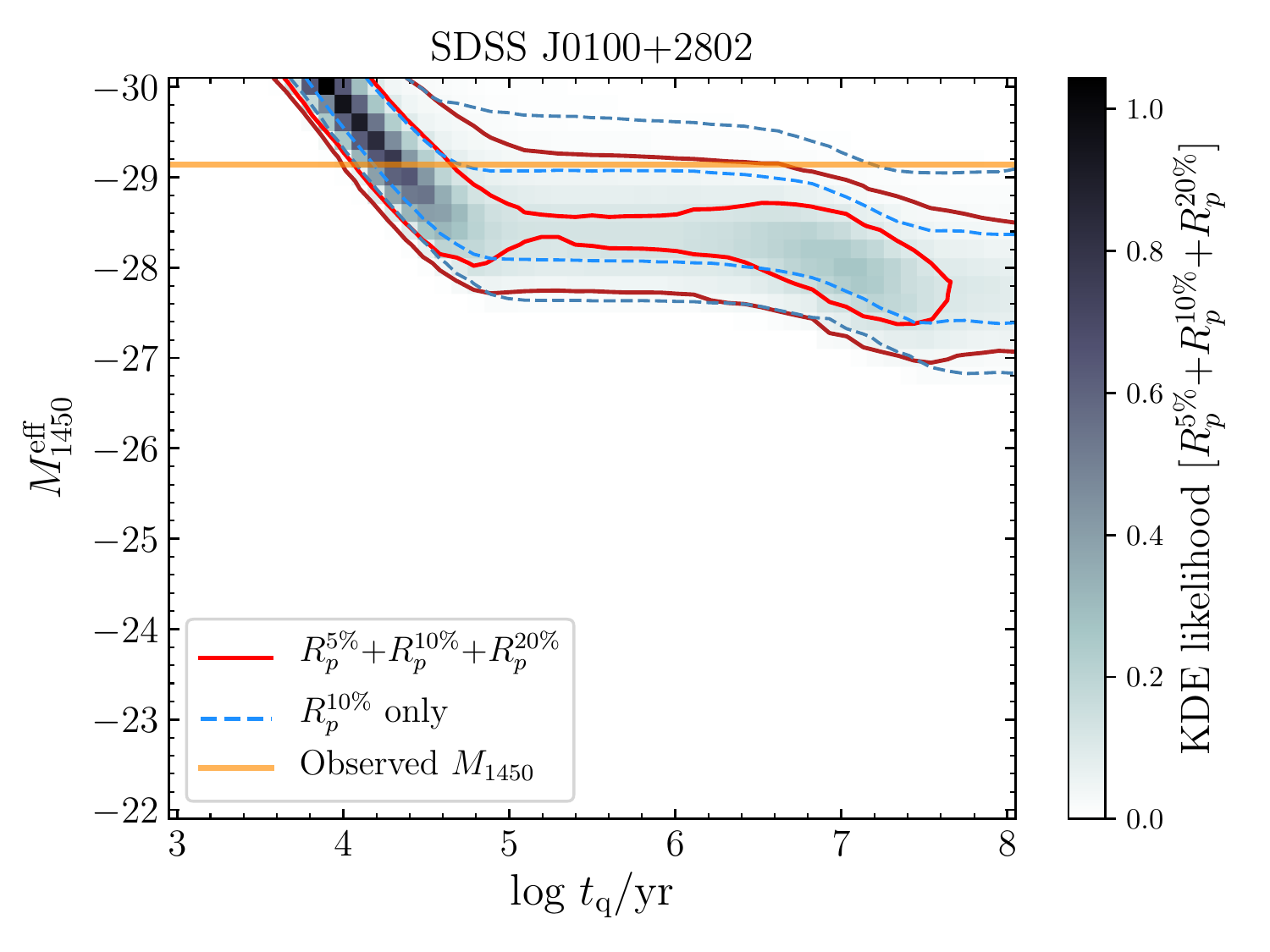}}
\resizebox{8cm}{!}{\includegraphics[trim={1.0em 1em 1.0em 1em},clip]{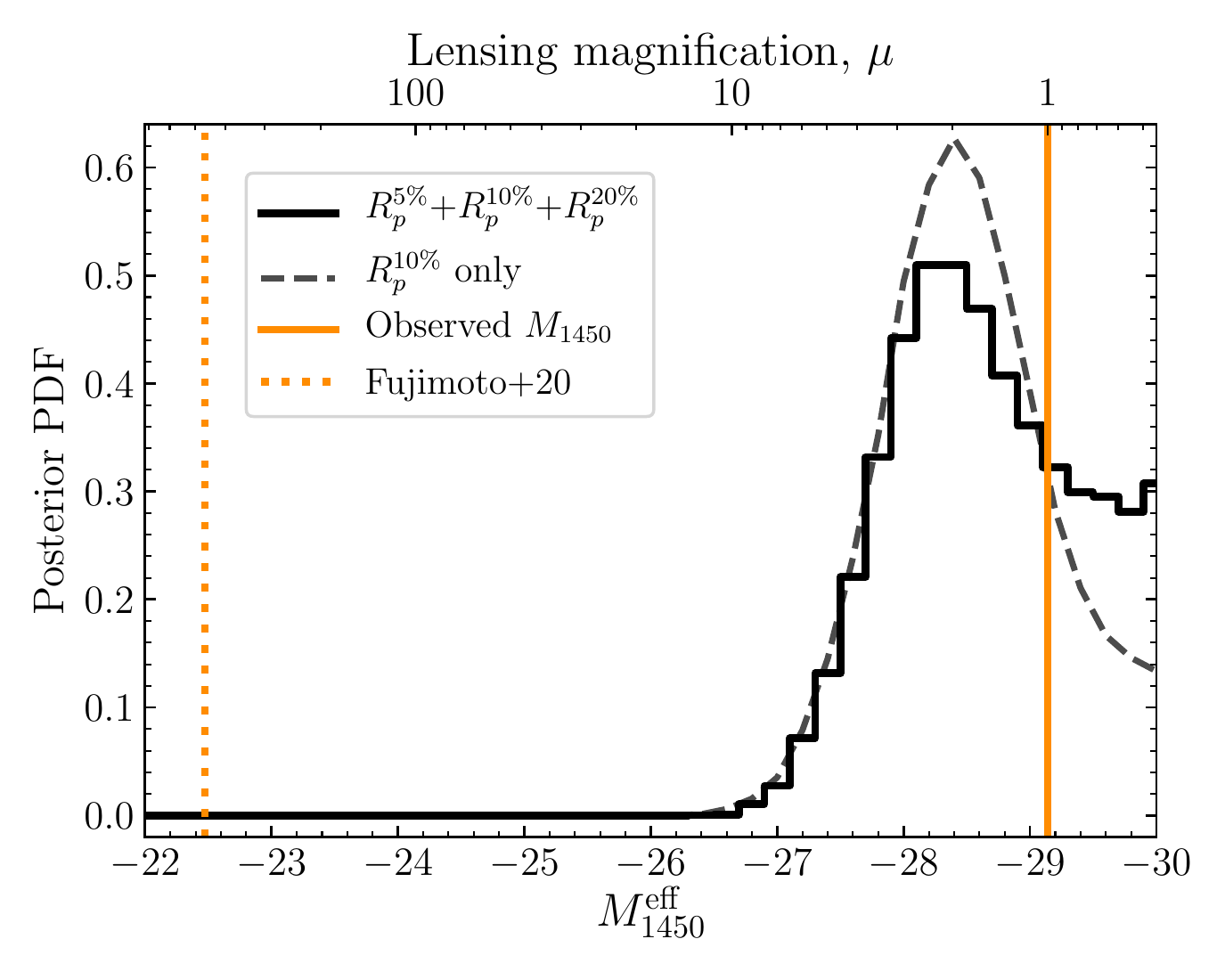}}
\end{center}
\caption{Analysis of the J0100+2802 proximity zone, similar to Figure~\ref{fig:j0439}. Left: Joint likelihoods of quasar lifetime and effective quasar absolute magnitude. Right: Marginalized posterior PDFs of $M_{1450}^{\rm eff}$, $\mu$. The vertical solid line shows the observed $M_{1450}$ of J0100+2802 ($\mu=1$), while the vertical dotted line corresponds to the magnification proposed by \citet{Fujimoto20} ($\mu=450$).}
\label{fig:j0100}
\end{figure*}

We first discuss the analysis of the lensed quasar J0439+1634. In the left panel of Figure~\ref{fig:j0439}, we show the 2D likelihood for the $R_p$ measurements from \S~\ref{sec:data}. The dashed curves show likelihood contours for $R_p^{10\%}$ alone, highlighting a near-perfect degeneracy between quasar lifetime and luminosity when only one measure of proximity zone size is employed. The solid curves and shading represent the joint likelihood of $\{R_p^{5\%},R_p^{10\%},R_p^{20\%}\}$, shifting the likelihood to fainter magnitudes and suppressing the likelihood at short quasar lifetimes. These differences are due to the relatively large $R_p^{5\%}$, which rules out short lifetimes due to the longer equilibration timescale at larger separations \citep{Davies20a}. Comparing to the orange line showing the apparent $M_{1450}=-29.75$, we see that non-lensed models are strongly disfavored.

In the right panel of Figure~\ref{fig:j0439} we show the posterior PDFs for $M_{1450}^{\rm eff}$ marginalized over lifetime and $\mu \equiv L_{\rm q,obs}/L_{\rm q,true}$ (i.e. the ratio between the apparent luminosity and the intrinsic, unlensed luminosity) using $R_p^{10\%}$-only (dashed) and $\{R_p^{5\%},R_p^{10\%},R_p^{20\%}\}$ (solid) as summary statistics, assuming a log-uniform prior on quasar lifetime (as in \citealt{Davies18b}). The posterior constraints on $\mu$ in both cases are consistent with the three lensing models proposed in \citet{Fan19}, shown as the vertical dotted lines -- unfortunately, we cannot strongly distinguish between them. Due to the degeneracy between $M_{1450}^{\rm eff}$ and $t_{\rm q}$ in the likelihood function, $R_p^{10\%}$ alone cannot fully rule out the unlensed hypothesis, but the peak still lies very close to the fiducial 3-image lens model of \citet{Fan19}. Including the additional $R_p$ values tightens the posterior PDF to a relatively narrow range: $\mu=28.0^{+18.4}_{-11.7}(^{+44.9}_{-18.3})$ at 68\% (95\%) credibility, and conclusively identifies the quasar as strongly lensed with $\mu>7.7$ at 99\% credibility.

We now turn to the nature of J0100+2802. In the left panel of Figure~\ref{fig:j0100}, we show the 2D likelihoods for $R_p^{10\%}$ and $\{R_p^{5\%},R_p^{10\%},R_p^{20\%}\}$ similar to Figure~\ref{fig:j0439}. Contrary to J0439+1634, the two likelihoods are quite similar, although using all three $R_p$ values modestly shifts the likelihood towards shorter quasar lifetime and brighter $M_{1450}^{\rm eff}$. In both cases the likelihoods intersect with the observed $M_{1450}=-29.14$. In the right panel of Figure~\ref{fig:j0100}, we show the marginalized posterior PDFs for $M_{1450}^{\rm eff}$ and the lensing magnification of J0100+2802. The posterior PDF implies $\mu<4.9$ ($\mu<7.0$) at 95\% (99\%) credibility. The $\mu\sim450$ model from \citet{Fujimoto20} is thus strongly disfavored by our analysis, even allowing for a factor of a few uncertainty in the lens model. 

\section{Discussion \& Conclusion} \label{sec:discuss}

Here we have shown that quantitative analysis of high-redshift quasar proximity zones can constrain their magnification by gravitational lensing. We first recovered the strong lensing of the known lensed quasar J0439+1634, suggesting that the proximity zone structure is a good probe of the intrinsic (ionizing) quasar luminosity. We then performed a similar analysis on the proximity zone spectrum of the hyperluminous quasar J0100+2802, and despite the smaller-than-expected proximity zone, we conclusively rule out the lensing magnification of $\sim450$ proposed by \citet{Fujimoto20}.


We note that a discrepancy between $M_{1450}^{\rm eff}$ and the observed $M_{1450}$ may not necessarily be due to lensing magnification. Namely, the ionizing SED of quasars is not perfectly known, and significant scatter is observed between the far-ultraviolet SEDs of individual quasars \citep{Telfer02}. Approximating the scatter in individual quasar power-law spectral indices from 1450{\AA} to 912{\AA} as a Gaussian with a $1\sigma$ dispersion of 0.5 (as a rough description of the distribution in \citealt{Telfer02}) suggests $1\sigma$ variations in the ionizing luminosity of $\sim0.1$ dex ($\sim0.25$ mag) from quasar to quasar at fixed $M_{1450}$. This scatter is small compared to the $R_p$-based constraints in this work due to the intrinsic ``noise'' from IGM density fluctuations, which are alone sufficient to match the typical scatter in observed $R_p$ \citep{Davies20a}.

Our posterior constraints on magnification assume a log-uniform prior over a wide range of quasar lifetime, $t_{\rm q}=10^3$--$10^8$\,yr. For J0100+2802, the most conservative constraint (i.e. with the highest magnification) comes from long quasar lifetimes of $t_{\rm q}\gtrsim10^7$\,yr. If we assume $t_{\rm q}=10^8$\,yr, our 95\% credibility constraint on the magnification of J0100+2802 relaxes somewhat to $\mu < 8$. However, recent analyses of reionization-epoch proximity zones \citep{Davies19} and \ion{He}{2} Ly$\alpha$ proximity zones \citep{Khrykin19} suggest that much shorter lifetimes on the order of $\sim10^6$\,yr are more common, and similar conclusions can be drawn from the fraction of quasars with short lifetimes \citep{Eilers17,Eilers20}. We also assumed a log-uniform prior over the lensing magnification; more informative priors could be chosen from lensing distribution models, which are roughly lognormal around $\mu\sim1$ with a long tail to higher values (e.g. \citealt{Hilbert08,Mason15lens,PL19}). However, such a prior would have to be balanced by the color and morphological selection biases against lensed objects.

With the exception of the serendipitous discovery of J0439+1634, direct searches for lensed quasars at $z>5$ with high-resolution imaging have failed to identify candidates \citep{Richards06lens,McGreer14}. We posit that high-redshift quasar proximity zones could be used to pre-select promising lensed candidates for followup imaging. The majority of $z\gtrsim6$ quasars have proximity zones which closely follow the expected sizes from our radiative transfer modeling \citep{Eilers17,Davies20a}, and for these quasars we do not expect there to be strong indications of lensing. A small subset of these quasars, however, have much smaller proximity zones than expected for their inferred luminosity \citep{Eilers17,Eilers20}, and for these quasars models with strong lensing could be preferred. Interestingly, a strongly lensed ($\mu>2$) fraction of $\sim10\%$ is consistent with the detection of J0439+1634 at $\mu\sim50$ in the observed sample of $z>6$ quasars in the lensing population models of \citet{PL19}, which is comparable to the fraction of quasars with small proximity zones \citep{Eilers20}.

Small proximity zone quasars which lack additional evidence supporting a youthful hypothesis (e.g. compact or non-existent halos of extended narrow-line emission, \citealt{Eilers18J1335,Farina19}) may represent the most likely candidates for lensed systems hiding in plain sight. That said, we have shown here that the young and lensed scenarios can potentially be distinguished using additional information contained within the full Ly$\alpha$ transmission profile. In future work, we will scour the known population of $z\gtrsim6$ quasars to identify those whose proximity zones imply strong lensing and warrant high spatial resolution investigation with e.g. HST or ALMA.

\acknowledgements

We would like to thank Sarah Bosman for useful discussions and advice regarding the flux calibration of the J0100+2802 VLT/X-SHOOTER spectrum. 

FW and ACE acknowledge support by NASA through the NASA Hubble Fellowship grants $\#$HF2-51448 and $\#$HF2-51434, respectively, awarded by the Space Telescope Science Institute, which is operated by the Association of Universities for Research in Astronomy, Inc., under NASA contract NAS5-26555.

Based on observations made with ESO Telescopes at the La Silla Paranal Observatory under program ID 096.A-0095.

\bibliographystyle{aasjournal}
 \newcommand{\noop}[1]{}

\end{document}